\newif{\ifjournal}
  \journaltrue

\ifjournal
  \documentclass[]{aa}
  \usepackage{graphicx,txfonts,amssymb,natbib}
  \renewcommand{\d}{\mathrm{d}}
  \authorrunning{M. Meneghetti et al.}
  \titlerunning{Strong lensing by cluster-sized haloes in dark-energy 
cosmologies}
\else
  \documentclass{paper}
\fi

\begin{document}

\title{Strong lensing by cluster-sized haloes in dark-energy 
cosmologies}
\ifjournal
  \author{Massimo Meneghetti\inst{1}, Matthias Bartelmann\inst{1},
    Klaus Dolag\inst{2}, Lauro Moscardini\inst{3}, Francesca
    Perrotta\inst{4,5,6}, \\  Carlo
    Baccigalupi\inst{4,5,6} and Giuseppe
    Tormen\inst{2} 
    \institute {$^1$ ITA, Universit\"at
    Heidelberg, Tiergartenstr.~15, D--69121 Heidelberg \\ 
    $^2$ Dipartimento di Astronomia,
    Universit\`a di Padova, Vicolo dell'Osservatorio 2, I--35120
    Padova \\ 
    $^3$
    Dipartimento di Astronomia, Universit\`a di Bologna, Via Ranzani
    1, I--40127, Bologna \\
    $^4$ SISSA/ISAS, Via Beirut 4, I--34014, Trieste \\ 
    $^5$ INFN, Sezione di Trieste, Via Valerio 2, I--34127 Trieste \\
    $^6$ Lawrence Berkeley National Laboratory, 1 Cyclotron Road,
    Berkeley, CA 94720, USA
}}
\else
  \author{{Massimo Meneghetti\inst{1}, Matthias Bartelmann\inst{1},
    Klaus Dolag\inst{2}, Lauro Moscardini\inst{3}, Francesca
    Perrotta\inst{4,5,6}, \\  Carlo
    Baccigalupi\inst{4,5,6} and Giuseppe
    Tormen\inst{2} 
    \institute {$^1$ ITA, Universit\"at
    Heidelberg, Tiergartenstr.~15, D--69121 Heidelberg \\ 
    $^2$ Dipartimento di Astronomia,
    Universit\`a di Padova, Vicolo dell'Osservatorio 2, I--35120
    Padova \\ 
    $^3$
    Dipartimento di Astronomia, Universit\`a di Bologna, Via Ranzani
    1, I--40127, Bologna \\
    $^4$ SISSA/ISAS, Via Beirut 4, I--34014, Trieste \\ 
    $^5$ INFN, Sezione di Trieste, Via Valerio 2, I--34127 Trieste \\
    $^6$ Lawrence Berkeley National Laboratory, 1 Cyclotron Road,
    Berkeley, CA 94720, USA}
\fi
\date{\emph{Astronomy \& Astrophysics, submitted}}

\newcommand{\abstext}
 {We study the efficiency of
  numerically simulated galaxy clusters for producing strong
  gravitational lensing events in different dark-energy cosmologies
  with constant and time-variable equation of state and we compare it
  with ``standard'' $\Lambda$CDM and OCDM models. Our main results are
  that: (1) the expected abundance of gravitational arcs with large
  length-to-width ratio depends on the equation of state of dark 
  energy at the epoch of formation of the halo; 
  (2) as found in previous studies, the strong lensing 
  cross section of galaxy clusters is very sensitive to dynamical
  processes like mergers, whose impact however is different for
  different cosmologies, being stronger for models in which halos are
  less concentrated. As expected, the largest differences in the
  lensing optical depth occur at moderate and high redshift.}

\ifjournal
  \abstract{\abstext}
\else
  \begin{abstract}\abstext\end{abstract}
\fi

\maketitle

\section{Introduction}

Evidence is mounting that the Universe is spatially flat, has low
matter density and is dominated by some form of dark energy, acting 
as a repulsive gravitational force and responsible for the present 
phase of accelerated cosmic expansion. 
\citep[e.g.][]{RI98.1,PE99.1,HA03.1,SP03.1,TE04.1,RI04.1}. 
The dark energy generalises the concept of cosmological constant,
admitting dynamics and fluctuations. Such a generalisation is
necessary attempting to reconcile its value with the known energy
scales of particle physics. The minimal extension of a cosmological
constant is a self-interacting scalar field
\citep[Quintessence, see][and references therein]{PE02.2}, already 
considered before the evidence for cosmic acceleration 
\citep{WE88.1,RA88.1}. 

Structure grows earlier in dark-energy than in cosmological-constant
models, which makes dark-matter haloes more concentrated
\citep{BA02.1,WE03.1,DO03.2}. Following earlier analytic work
\citep{BA03.1}, we study here with cluster models numerically
simulated in eight different cosmologies how earlier structure growth
and denser halo cores affect strong lensing by galaxy clusters. Our
motivation is three-fold. First, despite recent attempts
\citep{WA03.1,DA03.1}, it remains to be clarified whether the
strong-lensing efficiency of real clusters is reproduced by numerical
cluster models or not. Earlier studies found that the expected
strong-lensing efficiency of clusters in cosmological-constant models
appears to fall short by an order of magnitude of reproducing the
observed number of arcs
\citep{BA98.2}. We thus wish to quantify the effect on strong lensing of
the increased halo core density in dark-energy models. Second, earlier
structure growth may have interesting implications for strong cluster
lensing in view of the recently-detected clusters at redshifts near
unity in which strongly-lensed arcs were found
\citep{HA98.1,GL03.1,ZA03.1,LI04.1}. Third, we found in a recent study
with high time resolution \citep{TO04.1} that the efficiency of a
cluster for strong lensing can substantially be increased during
merger processes because of the enhanced tidal field and surface-mass
density during different phases of the merger process. Shifting
structure growth to earlier redshifts also allows major mergers at
higher redshifts and thus potentially much increased arc-formation
probabilities.

With these motivations in mind, we use the 17 cluster-sized haloes
produced for and described in \cite{DO03.2} and investigate their
strong-lensing properties. Section~2 describes the eight cosmological
models used, Sect.~3 the cluster sample, and Sect.~4 the lensing
simulations. Results are presented in Sect.~5 and summarised in
Sect.~6.

\section{Cosmological models \label{dem}}

We compare in this paper the efficiency for producing strong lensing
events of the sample of numerically simulated clusters previously
described in \cite{DO03.2}. We give here a short description of both
the cosmological models and the methods used for obtaining the cluster
sample and refer the reader to that paper for further details.

All clusters in the sample were simulated in different cosmologies
with constant and time-variable equation of state. These are an open
Cold Dark Matter (OCDM) and four spatially flat cosmological models,
namely a cosmological-constant ($\Lambda$CDM) model, a dark energy
model with constant equation of state (DECDM), and two quintessence
models with markedly different dynamical properties, one with inverse
power-law Ratra-Peebles potential \citep[RP, see][ and references
therein]{PE02.2} and one with SUGRA potential \citep[SUGRA, see][ and
references therein]{BR00.2}.  For the latter three we have used two
different normalisations of the perturbation power spectrum, to be
discussed later.

In all cases, the matter density parameter today is $\Omega_0=0.3$. In
the flat cosmologies, the remaining $70\%$ of the critical density is
assigned to the dark energy at present ($\Omega_{\Lambda}=0.7$. The
remaining cosmological parameters are $h=0.7$,
$\Omega_\mathrm{b}\,h^{2}=0.022$, Gaussian density fluctuations with
scale-invariant power spectrum, and no gravitational waves.

The normalisation of the perturbation power spectrum, which we set
conventionally through the \emph{rms} density fluctuation level
$\sigma_8$ within spheres of $8\,h^{-1}\,\mathrm{Mpc}$ radius, has
been chosen as $\sigma_8=0.9$ for all the cosmological models in order
to match the observed abundance of galaxy clusters. Since the power
spectrum normalisation is most important here, affecting the structure
statistics and therefore the lensing power, we also normalise the
DECDM, RP and SUGRA models to the observed level of anisotropies in
the Cosmic Microwave Background (CMB) \citep[e.g.][]{BE03.3}.  In this
case, the $\sigma_8$ is generally slightly smaller because of the
Integrated Sachs Wolfe (ISW) effect affecting the large scale CMB
anisotropies: the reason is the larger time interval in which the
cosmic equation of state changes in dark energy scenarios with respect
to the cosmological constant case. This enhances the dynamics of the
gravitational potentials and raises the ISW power, visible on the
large angle tail of CMB anisotropy power spectrum \citep[see][and
references therein]{BA02.1}.  We take $0.86$, $0.82$ and $0.76$ for
DECDM, RP and SUGRA, respectively. Note that the reduction is quite
important, since we selected our dark energy models in order to
highlight the differences among each other and the cosmological
constant. If the equation of state is smaller than the cases we
consider here, the ISW effect and the reduction to the power spectrum
normalisation are smaller.  These numbers as well as all the inputs
from the linear evolution of cosmological perturbations to the
N-body procedure described later are computed using our dark energy
oriented cosmological code, based on CMBfast \citep{SE96.1}. The code,
originally introduced by \citet{PE99.4}, describes the dark energy as
a scalar field, numerically integrating its background and linear
perturbation dynamics for the most relevant tracking quintessence
scenarios, capable to converge to the present phase of cosmic
acceleration from a wide range of initial conditions in the early
universe \citep{BA00.2}. This is performed in the general framework of
scalar-tensor cosmologies, in which the dark energy may be directly
linked with gravity \citep{PE00.1}. The outputs of this machinery have
been used for other work, allowing in particular to constrain the dark
energy equation of state with the modern CMB anisotropy observations
\citep{BA02.0}.

The dark energy cosmologies are characterised by their equation of
state parameter $w(z)$, describing the ratio between pressure and
density of the Dark Energy. For cosmic expansion to be accelerated
today, the $w$ parameter must be $<-1/3$. In the limit of constant
$w$, the dark energy density evolves with redshift proportionally to
$(1+z)^{3(1+w)}$. That is the case for both the $\Lambda$CDM model,
whose $w=-1$ yields a constant dark energy density, and the DECDM
model, which we choose to have $w=-0.6$. It is useful to note that
from the point of view of the Friedmann equation, the curvature term
in the OCDM model behaves as a dark energy component with constant
$w=-1/3$ \citep[see][ and references therein]{BA02.1}.

For the RP and SUGRA models, $w$ is time-dependent. The dark energy is
consistently described by means of the quintessence scalar field
$\phi$. The RP and SUGRA quintessence potentials are given by
\begin{equation}
  V_\mathrm{RP}=\frac{M^{4+\alpha_Q}}{\phi^{\alpha_Q}}\;,\quad
  V_\mathrm{SUGRA}=\frac{M^{4+\alpha_Q}}{\phi^{\alpha_Q}}
  \exp{(4\pi G\phi^2)}\;,
\label{RPSUGRA}
\end{equation}
respectively. The exponential in $V_\mathrm{SUGRA}$ reflects
super-gravity corrections \citep{BR00.2} and induces large variations
in the equation of state with respect to the RP case at the epoch of
structure formation, even if $w$ is the same at present in the two
models. \citet{DO03.2} found indeed that the cluster concentration
reflects the dark energy behaviour at the time of collapse, a promising
feature in order to reduce the degeneracy between different dark
energy models yielding the same amount of acceleration today. In this
work we check how this feature affects the statistics of strongly
lensed arcs in clusters. Specifically, the tracking RP and SUGRA
quintessence models considered here have $w_0=-0.83$ today, with
$\alpha_Q=-0.6$ and $\alpha_Q=-6.7$, respectively.  In the tracking
regime, this yields $w_{\rm SUGRA}\simeq -0.23$ and $w_{\rm RP}\simeq
-0.77$.

\section{Cluster sample}

The cluster models used in this study were simulated using a new
version of the cosmological code GADGET \citep{SP01.1}, specifically
adapted to simulate the formation and evolution of cosmic structures
in the dark energy models described earlier \citep[see][]{DO03.2}.

Each cluster in our sample was obtained using the so-called
re-simulation technique, which consists in re-simulating at higher
resolution a patch of a pre-existing large-scale cosmological
simulation \citep{TO97.2}. For this work, we selected as a parent simulation an
N-body run with $512^3$ particles in a box of
$479\,h^{-1}\,\mathrm{kpc}$ \citep{YO01.1,JE01.1}. Its background
cosmological model is spatially flat with $\Omega_\mathrm{m,0}=0.3$
and $\Omega_\Lambda=0.7$ at the final epoch, identified with redshift
zero. The particle mass was $6.8\times10^{10}\,h^{-1}\,M_\odot$, and
the gravitational softening was chosen as
$30\,h^{-1}\,\mathrm{kpc}$. From the output of this simulation at
$z=0$, we selected ten spherical regions of radius between $5$ and
$10\,h^{-1}\,\mathrm{Mpc}$, each containing either one or a pair of
dark matter haloes with mass exceeding $3\times10^{14}\,h^{-1}\,M_\odot$. We
thus obtain a sample of 17 cluster-sized objects.

Each of these regions was re-sampled to build new initial conditions
with on average $10^6$ dark-matter particles. The initial conditions
from the $\Lambda$CDM cosmology were adapted to all the dark energy
cosmologies studied here as described in \citet{DO03.2}. The mass
resolution of the re-simulation ranges from $2\times10^9$ to
$6\times10^9\,h^{-1}\,M_\odot$ per dark-matter particle so as to have
each cluster consisting of approximately the same number of
particles. The gravitational softening is reduced to a
$5\,h^{-1}\,\mathrm{kpc}$ cubic spline smoothing for all particles in
the high-resolution region.

We found that our 17 clusters contain on average $N_V\approx200,000$
dark matter particles within their virial radii. The corresponding
virial masses range between $M_V=3.1\times10^{14}$ to
$1.7\times10^{15}\,h^{-1}\,M_\odot$.

\section{Lensing simulations}

Ray-tracing simulations are then carried out with each of the 17
clusters in all eight cosmological models. The technique used in this
study was described in detail in several earlier papers
(e.g.~\citealt{BA98.2,ME00.1}).

For all clusters in each cosmological model, we produce $N_{\rm
snap}=52$ snapshots at different redshift between 0 and 1. These are
equidistant in time, leading to a time resolution of $\sim100\,$Myr,
which allows the effects of dynamical processes on the cluster cross
sections for strong lensing to be resolved, as discussed in
\citet{TO04.1}.

For each snapshot, we select a cube of $3\,h^{-1}$Mpc comoving side
length, centred on the halo centre and containing the high-density
region of the cluster. The particles in this cube are used for
producing a three-dimensional density field, by interpolating their
position on a grid of $256^3$ cells using the {\em Triangular Shaped
Cloud} method \citep{HO88.1}. Then, we project the three-dimensional
density field along the coordinate axes, obtaining three surface
density maps $\Sigma_{i,j}$, used as lens planes in the following
lensing simulations. The total number of lensing simulations is
therefore $17\times52\times3\times8=21216$, requiring approximately
the same number of computation hours on an IBM-SP4 cluster located at
the computer centre of the Max Planck Society in Garching.

The lensing simulations are performed by tracing a bundle of $2048
\times 2048$ light rays through a regular grid, covering the central
quarter of the lens plane. This choice is driven by the necessity of
studying in detail the central region of the clusters, where critical
curves form, taking into account the contribution from the surrounding
mass distribution to the deflection angle of each ray.

Deflection angles on the ray grid are computed using the method
described in \citet{ME00.1}. We first define a grid of $128\times128$
``test'' rays, for each of which the deflection angle is calculated by
directly summing the contributions from all cells on the surface
density map $\Sigma_{i,j}$,
\begin{equation}
  \vec \alpha_{h,k}=\frac{4G}{c^2}\sum_{i,j} \Sigma_{i,j} A
  \frac{\vec x_{h,k}-\vec x_{i,j}}{|\vec x_{h,k}-\vec x_{i,j}|^2}\;,
\end{equation}  
where $A$ is the area of one pixel on the surface density map and
$\vec x_{h,k}$ and $\vec x_{i,j}$ are the positions on the lens plane
of the ``test'' ray ($h,k$) and of the surface density element
($i,j$). Following \cite{WA98.2}, we avoid the divergence when the
distance between a light ray and the density grid-point is zero by
shifting the ``test'' ray grid by half-cells in both directions with
respect to the grid on which the surface density is given. We then
determine the deflection angle of each of the $2048\times2048$ light
rays by bi-cubic interpolation between the four nearest test rays.

The position $\vec y$ of each ray on the source plane is calculated by
applying the lens equation. If $\vec y$ and $\vec x$ are the angular
positions of source and image from an arbitrarily defined optical axis
passing through the observer and perpendicular to the lens and source
planes, this is written as
\begin{equation}
  \vec y = \vec	x -\frac{D_{\rm ls}}{D_{\rm s}}\vec \alpha(\vec x)\;,
\end{equation}
where $D_{\rm ls}$ and $D_{\rm s}$ are the angular diameter distances
between the lens and the source plane, and between the observer and
the source plane, respectively.

Then, a large number of sources is distributed on the source plane. We
place this plane at redshift $z_\mathrm{s}=1$. Keeping all sources at
the same redshift is an approximation justified for the purposes of
the present case study, but the recent detections of arcs in
high-redshift clusters \citep{ZA03.1,GL03.1} indicate that more
detailed simulations will have to account for a wide source redshift
distribution.

The sources are elliptical with axis ratios randomly drawn from
$[0.5,1]$. Their equivalent diameter (the diameter of the circle
enclosing the same area of the source) is $r_\mathrm{e}=1''$. These
are distributed on a region on the source plane corresponding to one
quarter of the field of view where rays are traced. As in our earlier
studies, we adopt an adaptive refinement technique when placing
sources on their plane. We first start with a coarse distribution of
$32\times32$ sources and then increase the source number density
towards the high-magnification regions of the source plane by adding
sources on sub-grids whose resolution is increased towards the lens
caustics. This increases the probability of producing long arcs and
thus the numerical efficiency of the method. In order to compensate
for this artificial source-density enhancement, we assign a
statistical weight to each image for the following statistical
analysis which is proportional to the area of the sub-grid cell on
which the source was placed.

By collecting rays whose positions on the source plane lay within any
single source, we reconstruct the images of background galaxies and
measure their length and width. Our technique for image detection and
classification was described in detail by \cite{BA94.1} and adopted by
\cite{BA98.2}, \cite{ME00.1,ME01.1,ME03.2,ME03.1} and \cite{TO04.1}. It
results in a catalogue of simulated images which is subsequently
analysed statistically.

\section{Results}

In this Section, we discuss how the lensing properties of the clusters
in our sample differ in the different cosmological models.

\subsection{Critical curves and caustics}

\begin{figure*}
  \includegraphics[width=\hsize]{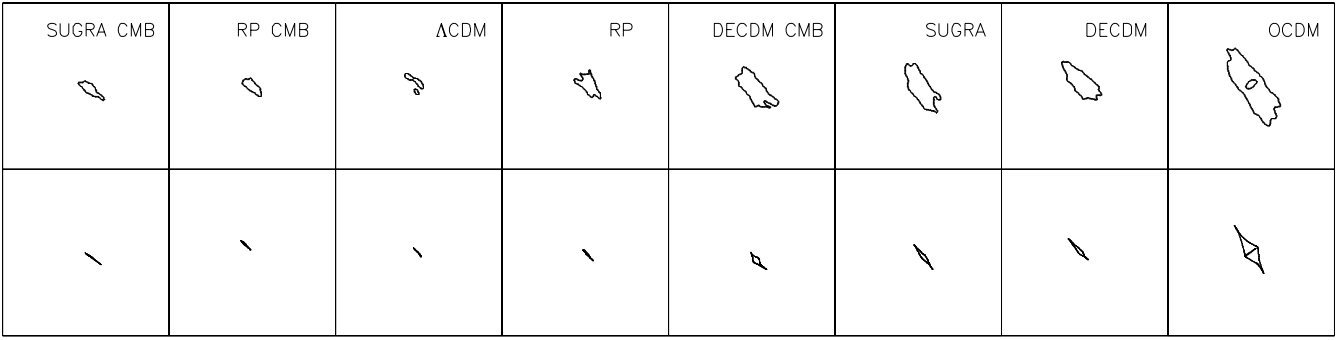}
\caption{Critical curves (upper panels) and caustics (bottom panels)
  of one of the clusters in our sample in all eight cosmological
  models. The side length of each panel is $50''$}
\label{fig:critcaust}
\end{figure*}

We start by considering the critical curves, along which strongly
magnified and highly distorted images form. These are the lines on the
lens plane where the determinant of the Jacobian matrix of the lens
mapping,
\begin{equation}
  A_{hk}(\vec{x}) \equiv \frac{\partial{y}_h}{\partial{x}_k} =
  \delta_{hk}-\frac{\partial\alpha_h}{\partial x_k}\;,
\label{jacobian}
\end{equation}   
is zero. Since the local magnification is the inverse of the Jacobian
determinant, the critical points are ideally characterised by infinite
magnification.

Through the lens equation, critical lines are mapped onto the caustics
on the source plane. Sources lying close to the caustics therefore
have images strongly magnified in the radial or tangential directions,
i.e.~gravitational arcs. The probability of one cluster to produce
highly distorted images is then related to the size and the shape of
the critical curves and of the caustics.

In Fig.\ref{fig:critcaust} we show some examples of how the critical
curves (and of the corresponding caustics) of one of the clusters in
our sample change among different cosmological models. The halo has a
virial mass ranging between $4.3\times 10^{14}\,h^{-1}\,M_\odot$ and
$5.2\times 10^{14}\,h^{-1}\,M_\odot$ at redshift $z \sim
0.3$. Considering models characterised by the same normalisation of
the power spectrum of the primordial density perturbations, it is
clearly seen that the two models with dynamical dark energy, i.e.~the
RP and SUGRA models, and the DECDM model with constant equation of
state $w=-0.6$, interpolate between the $\Lambda$CDM and the OCDM
model. In particular, the critical curves in the SUGRA model have
sizes comparable to those in the DECDM model.

However, it is important to note that the size of the critical curves 
and of the caustics in dark energy models may be smaller when the CMB 
normalisation is used, because of the enhanced ISW effect with respect 
to the $\Lambda$CDM case. The larger the 
required reduction of $\sigma_8$ is, the more the critical curves
shrink; in the case considered here, the critical curves 
tend to have sizes comparable to those in the $\Lambda$CDM
model or smaller. 

\subsection{Lensing cross sections}

In earlier work, the length-to-width ratio of gravitational arcs was
found to be very sensitive to many intrinsic properties of the lenses
which depend on cosmology. In fact, numerical simulations showed that
order-of-magnitude differences are expected in the number of arcs on
the whole sky with length-to-width ratio exceeding a given threshold
in different cosmological models \citep{BA94.1,BA95.1,BA98.2}. In
particular, ten times more arcs are expected in an OCDM than in a
$\Lambda$CDM model. Modelling clusters using the NFW density profile,
which depends on cosmology in contrast to the singular isothermal
profile, such a high sensitivity of the abundance of long and thin
arcs on cosmology has been confirmed even using an analytic approach
\citep{ME03.1,BA03.1}.

Therefore, aiming at evaluating the differences between the
strong-lensing efficiency of clusters in different cosmological models
with dark energy, we focus on the statistical distributions of the arc
length-to-width ratios.

The efficiency of one galaxy cluster for producing arcs with a given
property can be quantified by means of its lensing cross section. By
definition, this is the area on the source plane where a source must
be placed in order to be imaged as an arc with that property. As
explained in the previous sections, we artificially increase the
density of background sources close to the cluster caustics by
adopting adaptive grid refinement. Each source is taken to represent a
fraction of the source plane. We assign to each source and all of its
images a statistical weight $w$ which is inversely proportional to the
squared resolution of the sub-grid on which the corresponding source
was placed. The cells of the sub-grid with the highest resolution have
area $A$, and the sources placed on its grid points are given a
statistical weight of unity. The absolute lensing cross sections are
then determined by counting the statistical weights of the sources
whose images have a length-to-width ratio exceeding a threshold
$(L/W)_{\rm min}$. If a source has multiple images with
$(L/W)\ge(L/W)_{\rm min}$, we multiply its statistical weight by the
number of such images. Therefore, the lensing cross section is
\begin{equation}
\label{equation:ncross}
  \sigma_{(L/W)_{\rm min}}=A\,\sum_i\,W_i w_i n_i\;,
\end{equation}
where $W_i$ is unity if the $i$-th source has images with
$(L/W)\ge(L/W)_{\rm min}$ and zero otherwise, $n_i$ is the number of
images of the $i$-th source satisfying the required condition, and
$w_i$ is the statistical weight of the source.

\subsubsection{Averaged cross sections}

\begin{figure*}
  \includegraphics[width=0.33\hsize]{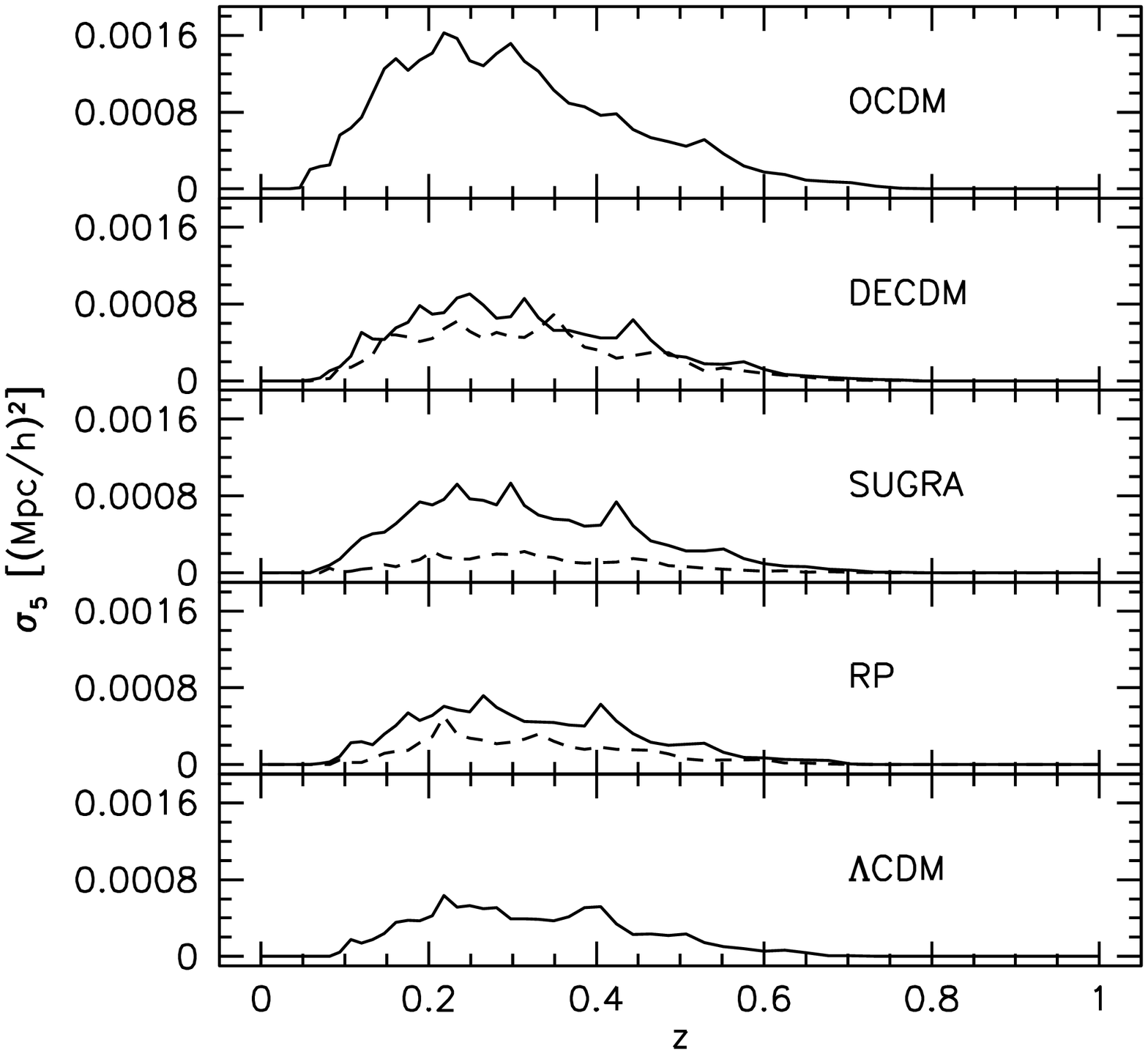}
  \includegraphics[width=0.33\hsize]{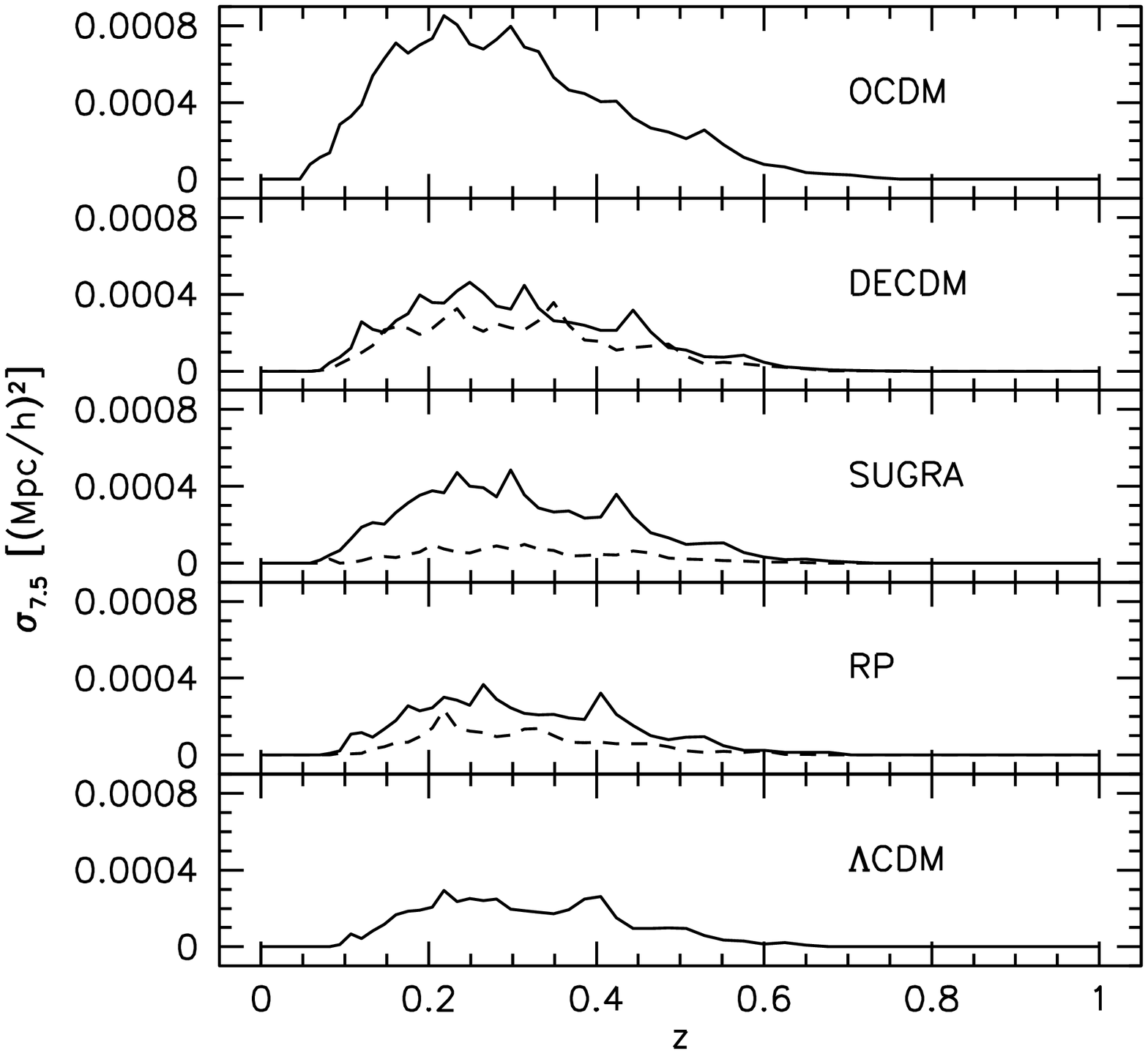}
  \includegraphics[width=0.33\hsize]{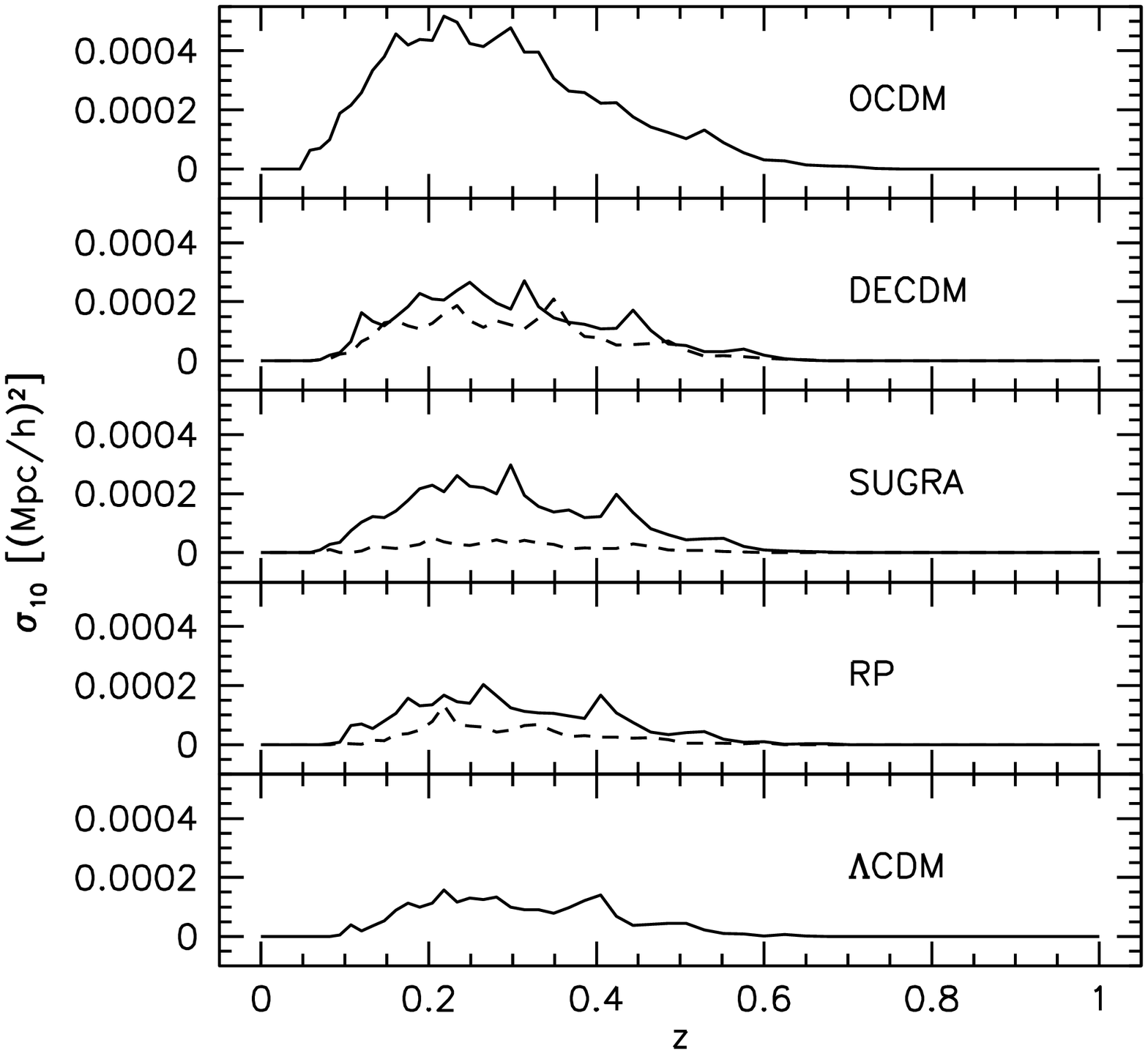}
\caption{Averaged lensing cross sections for arcs with length-to-width
  ratio larger than 5 (left panel), 7.5 (central panel) and 10 (right
  panel) of our cluster sample as function of the lens
  redshift. Sources are kept at redshift $z_{\rm s}=1$. Different
  panels refer to different cosmological models. Solid curves show the
  results for cosmologies with $\sigma_8=0.9$. Dashed curves show the
  correspondent results when $\sigma_8$ is reduced for taking into
  account the increasing Sachs-Wolfe effect affecting the large scales
  in the CMB.}
\label{fig:crosstan}
\end{figure*}

We now discuss the averaged cross sections of our cluster sample. In
Fig.~\ref{fig:crosstan} we show the lensing cross sections for long
and thin arcs, averaged over all 17 clusters, as function of the lens
redshift for all eight cosmological models. For each cluster, we use
its three projections along the coordinate axes for measuring the
cross section. Therefore, the curves displayed in
Fig.~\ref{fig:crosstan} result from averaging over 51 different curves
in each cosmological model.

As expected, the lensing cross sections reflect the differences in the
concentration of dark matter halos in different cosmological
models. This has been fully discussed in our previous paper
\citep{DO03.2}. It is shown there that the cluster concentration
depends on the dark energy equation of state at the cluster formation
redshift $z_\mathrm{coll}$ through the linear growth factor
$D_+(z_\mathrm{coll})$. Assuming the same normalisation of the power
spectrum, the lensing cross sections for the OCDM and the $\Lambda$CDM
models differ by roughly a factor of four, independent of the minimal
length-to-width ratio of the arcs, and the cross sections for the
other cosmological models with dark energy interpolate between
them. Despite the equation of state of the dark energy today is the
same for the RP and SUGRA models, their lensing cross sections differ
substantially at higher redshift.

For models with the CMB normalisation of the power spectrum, we find
lensing cross sections smaller by even more than one order of
magnitude compared to the OCDM model. In fact, when the normalisation
is reduced because of the Integrated Sachs Wolfe (ISW) effect
affecting the large-scale CMB anisotropies in the cosmologies we
consider \citep{BA03.1}, not only the formation epoch of our simulated
clusters is delayed, but also their evolution up to redshift zero is
changed. For example, clusters in the RP and in the SUGRA models have
masses at redshift zero which are smaller by roughly $20\%$ and $30\%$
respectively compared to the $\Lambda$CDM model.

The cluster sample is still too small for the averaged cross section
to be a smooth function of redshift. In fact, the curves exhibit
strong peaks which are caused by merger events arising in single
clusters. It has recently been shown that during such events, on
timescales of some hundreds of Myr, the cluster efficiency for strong
lensing is strongly enhanced \citep{TO04.1}, due to the combined
effects of the increasing shear and convergence. We will discuss the
properties of these peaks in the next subsection, when the lensing
cross sections of one individual cluster will be shown.

\subsubsection{Individual clusters}

\begin{figure*}
  \includegraphics[width=.33\hsize]{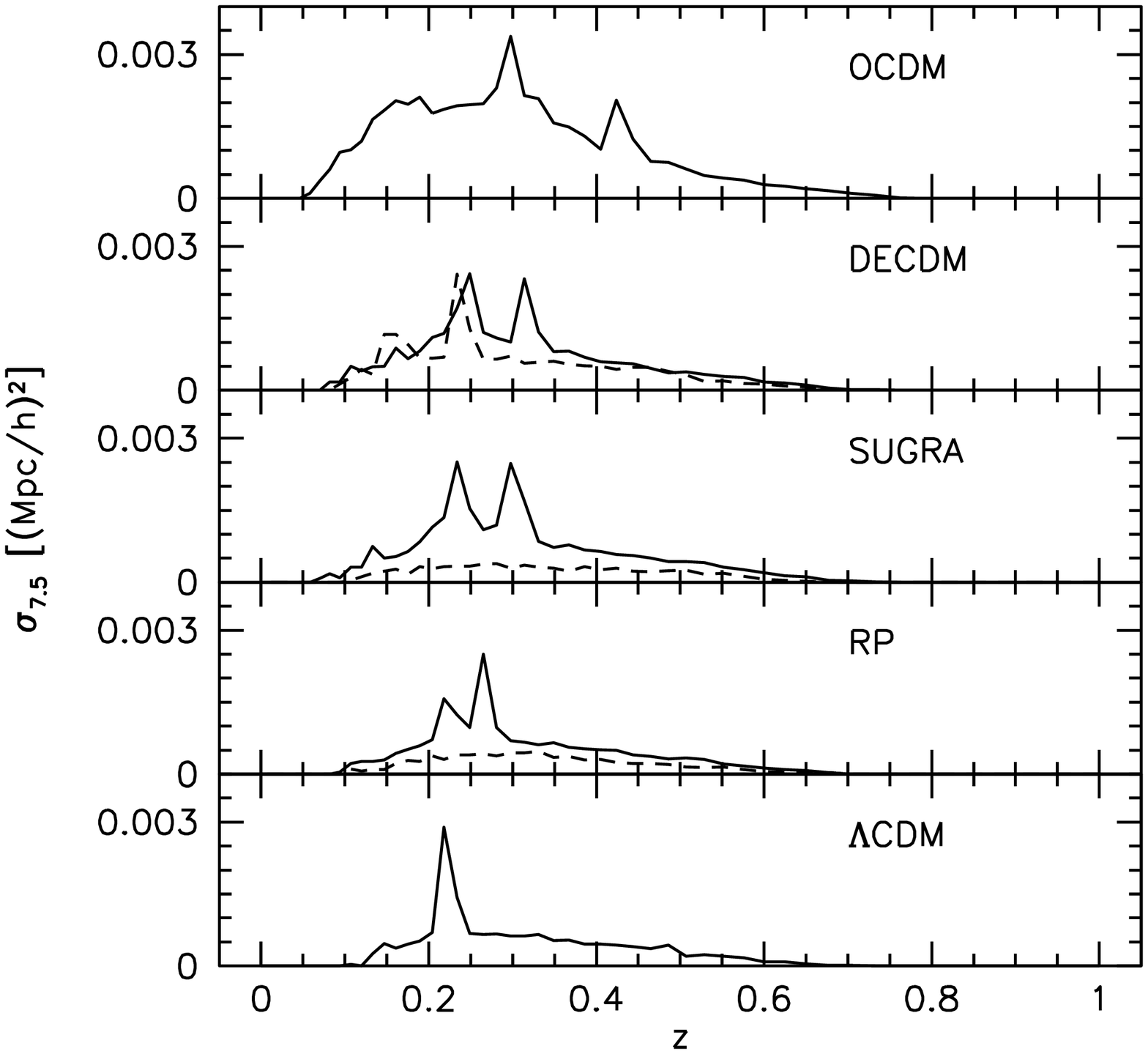}
  \includegraphics[width=.33\hsize]{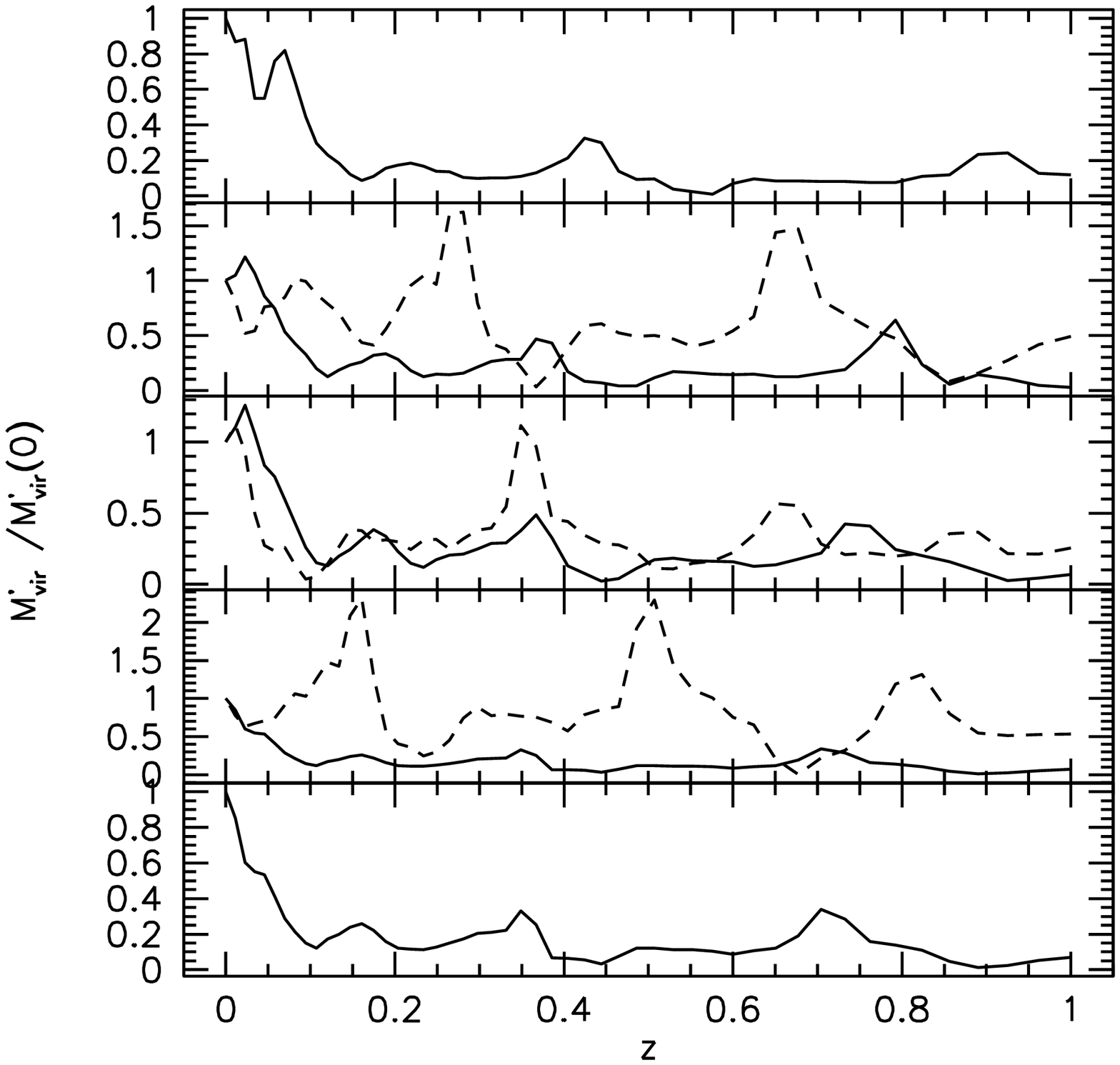}
  \includegraphics[width=.33\hsize]{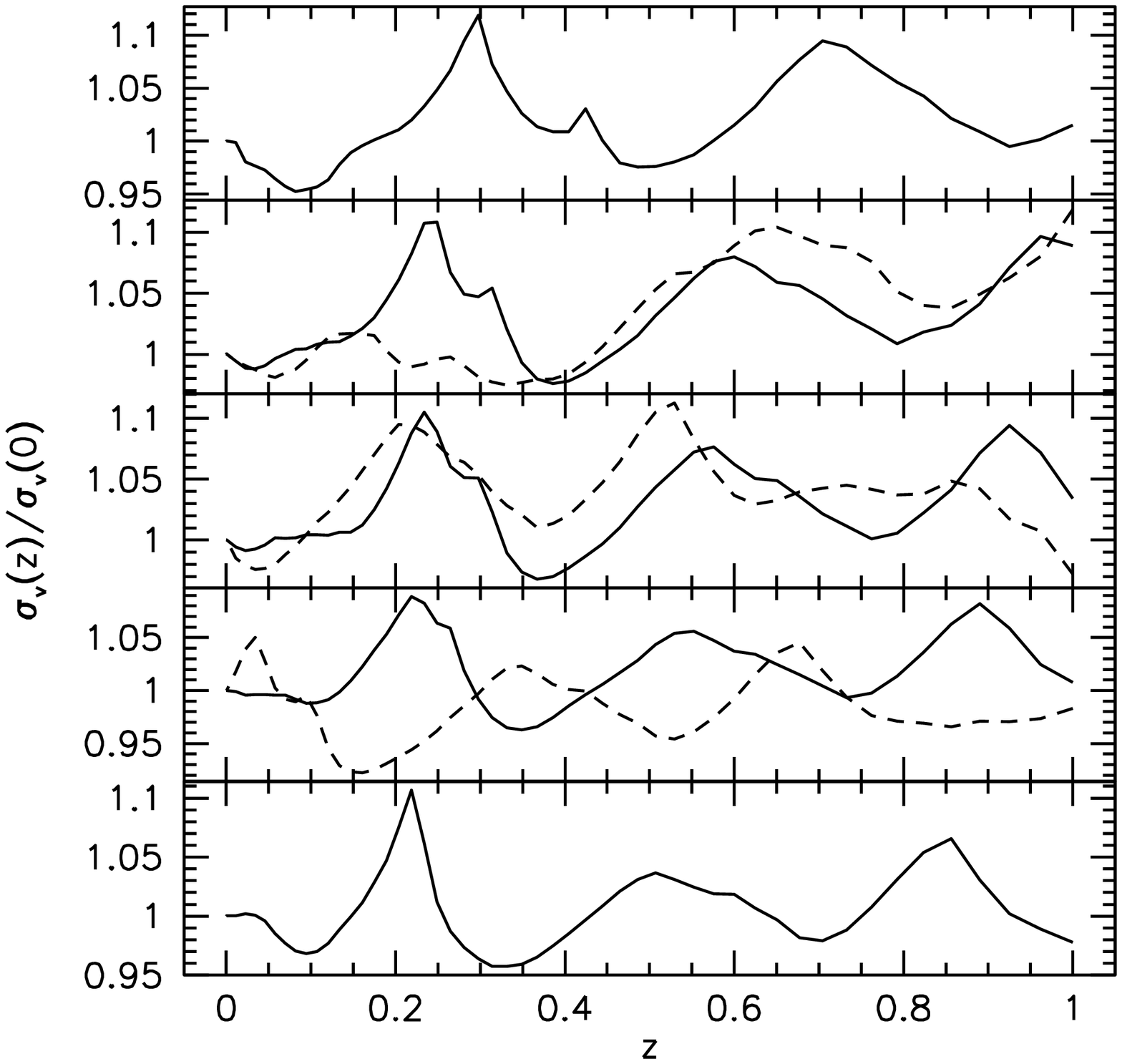}
\caption{Left panel: lensing cross sections for arcs with
  length-to-width ratio larger than 7.5 of one of the most massive
  clusters in our sample as a function of the lens redshift in different
  cosmological models. As in Fig.~\ref{fig:crosstan}, sources are kept
  at redshift $z_{\rm s}=1$. Solid curves show the results for
  cosmologies with $\sigma_8=0.9$. Dashed curves show the
  corresponding results for $\sigma_8$ reduced according to CMB
  normalisation. Central and right panels: redshift evolution of the
  derivative of the mass with respect to redshift and of the velocity
  dispersion of the same cluster as in the left panel. Solid and
  dashed lines refer to the different normalisations of the power
  spectrum.}
\label{fig:g10csec}
\end{figure*}

In Fig.~\ref{fig:crosstan}, several peaks can be found in at least all
the curves obtained in cosmological models with the same normalisation
of the power spectrum. Indeed, in this case, we are comparing the same
clusters in different cosmological models and therefore the same
dynamical processes cause the cluster lensing efficiencies to
peak. However the position and the amplitude of the peaks is strongly
dependent on the cosmological model. This becomes clearer by looking at
the lensing cross sections of one individual cluster.

In the left panel of Fig.\ref{fig:g10csec} we show the lensing cross
section for arcs with length-to-width ratio larger than 7.5 of one of
the most massive clusters in our sample ($\sim
10^{15}\,h^{-1}\,M_\odot$). Solid curves refer to models with
$\sigma_8=0.9$. Since the cluster forms at different epochs in
different cosmological models, the same merger events occur at
different redshifts, and consequently the peaks are shifted in
time. Initial conditions of our clusters are built such that single
halos reach the same evolutionary state at redshift zero in all
cosmological models. Therefore, the delay between dynamical processes
in different cosmologies shrinks as the redshift decreases.

The offset between the merger histories is clear from the central and
the right panel of Fig.~\ref{fig:g10csec}. In the central panel, the
redshift evolution of the redshift derivative of the halo mass within
the virial region, defined as the region enclosing 200 times the
\emph{mean} background density, is shown. The mass grows towards
redshift zero as the cluster accretes material from its
surroundings. Sudden peaks in the curves indicate that large clumps of
matter are entering into the virial region. As discussed in
\cite{TO04.1}, the typical time-scale for an infalling substructure to
reach the peri-centre of its orbit is then $\sim1$ Gyr, when the
passage of the substructure produces a peak in the particle velocity
dispersion, as shown in the right panel of Fig.~\ref{fig:g10csec}. All
the most pronounced peaks in the curves representing the lensing cross
sections correspond to some peak in the velocity dispersion, which
demonstrates our previous assertion about the relation between merger
events and maxima in the cluster lensing efficiency.

By comparing the left and the right panels of Fig.~\ref{fig:g10csec},
we note that we cannot find for all the peaks in the velocity dispersion
a corresponding peak in the lensing cross sections. The reason is
that peaks at high and low redshift are suppressed by the geometrical
lensing efficiency. Indeed, the lensing cross sections drop when the
lenses are too close to the observer or to the sources.

It is remarkable that even if the lens is intrinsically not very
efficient for lensing, the passage of massive substructures close to
the cluster centre can strongly enhance its ability to produce
gravitational arcs. For example, in the $\Lambda$CDM model the lensing
cross section is larger by roughly one order of magnitude while a
merger is occurring. Similar increments in the amplitude of the
lensing cross section have been found by \cite{TO04.1}. This suggests
that lensing events are likely to be transient events, characterising
peculiar epochs during cluster evolution.

When the CMB normalisation of the power spectrum is adopted, the
comparison between the same clusters in different cosmological models
is less straightforward. Indeed, changing $\sigma_8$ strongly affects
the merging history of the cluster, as shown by the dotted lines in
Fig.~\ref{fig:g10csec}.

\begin{figure*}
  \includegraphics[width=.5\hsize]{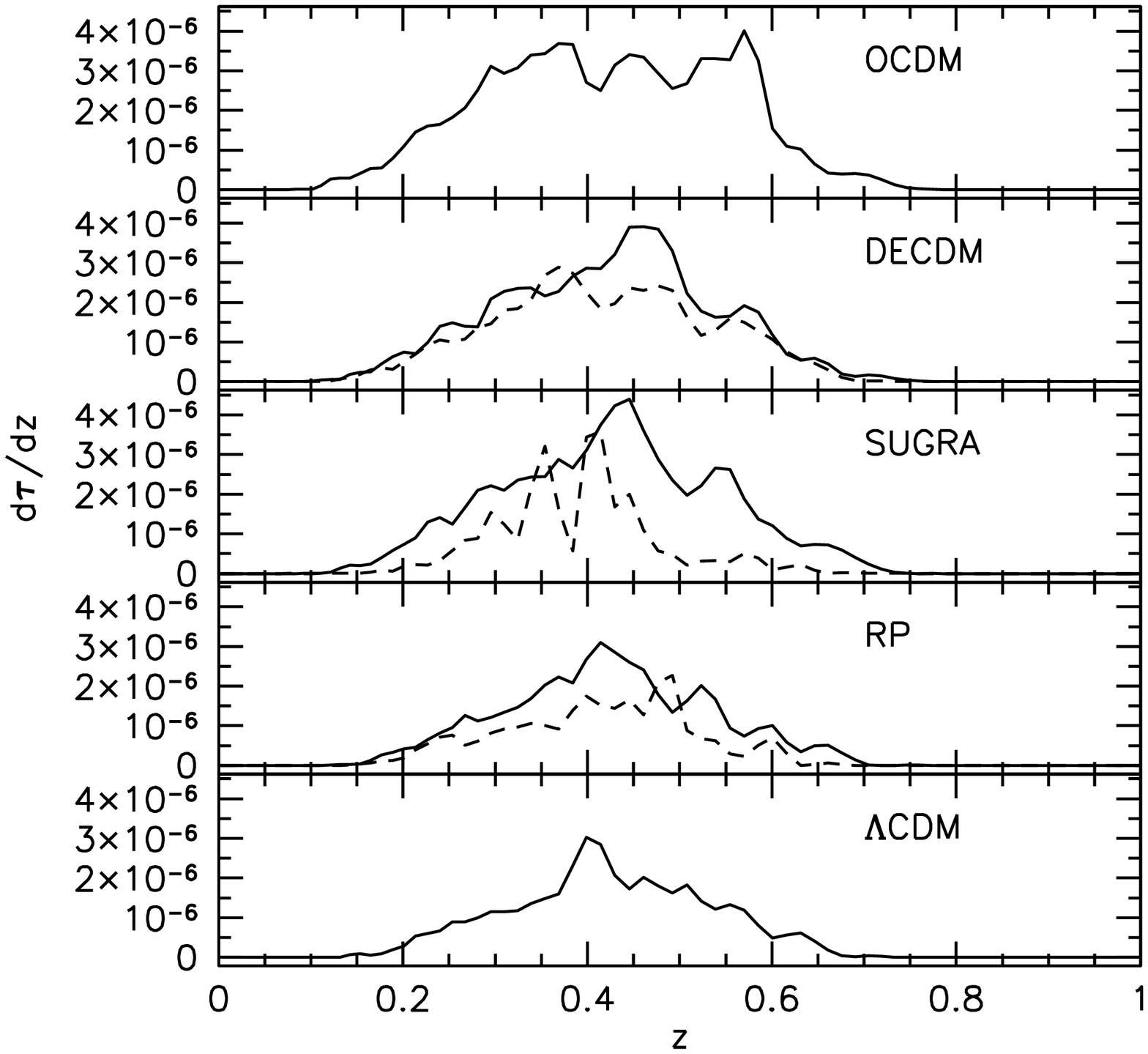}
  \includegraphics[width=.5\hsize]{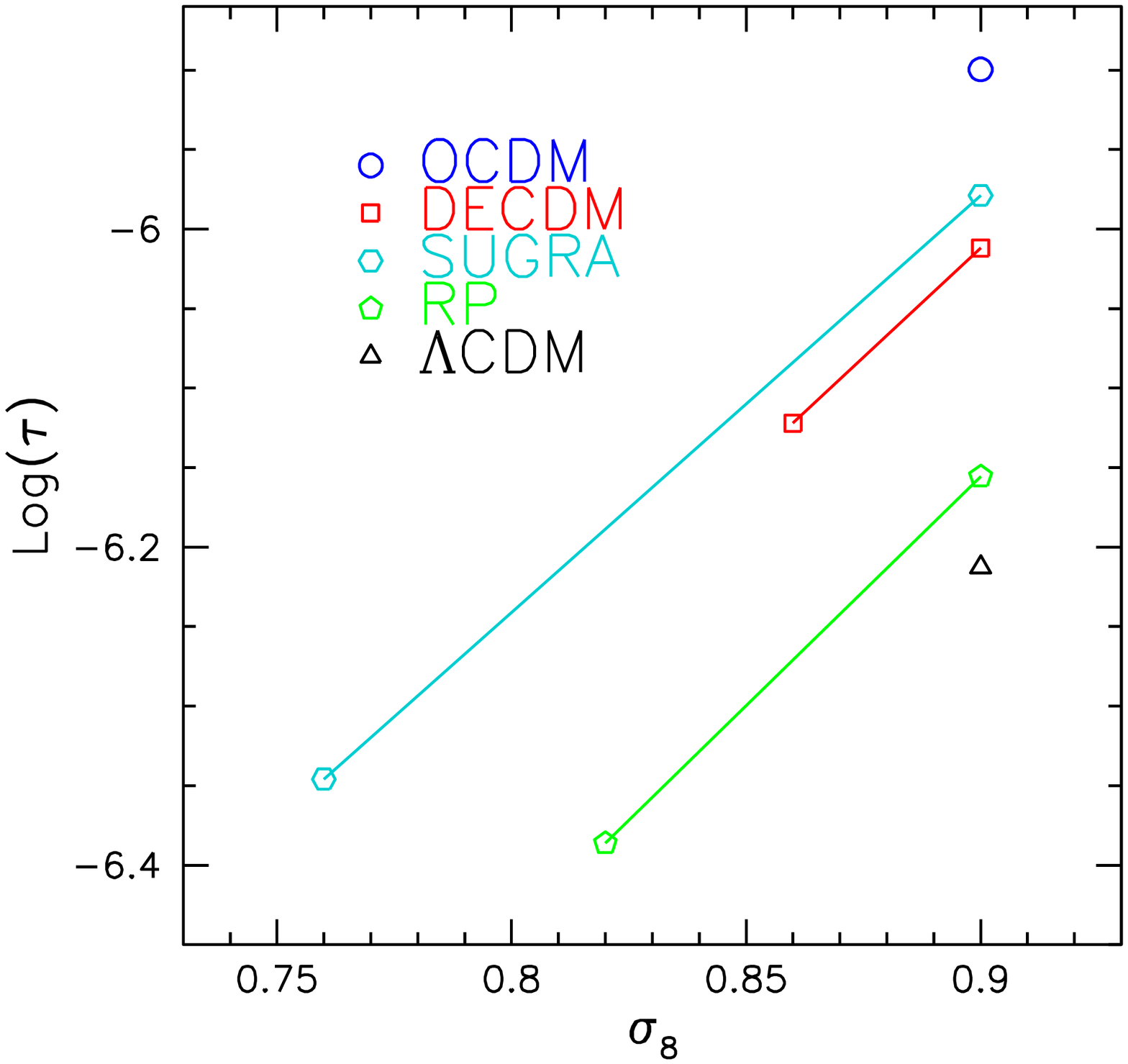}
\caption{Left panel: differential optical depth for arcs with
  $L/W>7.5$ and sources at $z_{\rm s}=1$ as a function of the lens
  redshift. Solid curves show the results for cosmologies with
  $\sigma_8=0.9$. Dashed curves show the corresponding results when
  $\sigma_8$ is reduced according to CMB normalisation. Right panel:
  optical depth for arcs with $L/W>7.5$ and sources at $z_{\rm s}=1$
  as a function of the normalisation of the power spectrum.}
\label{fig:optd}
\end{figure*}

We also note that in both the RP and the SUGRA models with CMB
normalisation, even when the cluster appears to be dynamically
perturbed, the lens still does not become particularly efficient for
strong lensing. We verified that in these cases, the substructures
producing the peaks in the velocity dispersion are orbiting at large
distance from the cluster centre. Indeed, the merger events producing the
peaks in the lensing cross sections in the cluster-normalised models
are here delayed in time due to the reduced $\sigma_8$.  

\subsection{Optical depth}

We can now estimate the probability for arc formation of a population
of clusters. Given a mass function, $\d n(M,z)/\d z$, assuming that
the typical lensing cross sections for lenses of mass $M$ are well
represented by the lensing cross sections of our sample $\sigma(M,z)$,
the optical depth $\tau(z_{\rm s})$ is defined as the sum of the
lensing cross sections of each lens between the observer and the
sources, divided by the area of the source plane. Following this
definition,
\begin{equation}
  \tau(z_\mathrm{s})=\frac{1}{4 \pi D_{\rm s}^2}\int_0^{z_\mathrm{s}}\d z\,
  (1+z)^3\left|\frac{\d V}{\d z}\right|\,
  \int_{0}^\infty\d M\,
  \frac{\d n}{\d M}\,\sigma(M,z)\;,
\label{equation:taudef}
\end{equation}
where $z_\mathrm{s}$ is the source redshift, $V$ is the cosmic volume,
and the factor $(1+z)^3$ accounts for the fact that the cluster number
density is defined per comoving volume. The former equation applies
under the assumption that lensing cross sections do not overlap.

Since we do not know $\sigma(M,z)$ for a continuous range of masses,
but only for a discrete cluster sample $M_i$, we calculate the optical
depth as follows:
\begin{eqnarray}
  \tau(z_\mathrm{s}) &=&\frac{1}{4 \pi D_{\rm s}^2}\int_0^{z_\mathrm{s}}\d z\,
  (1+z)^3\left|\frac{\d V}{\d z}\right|\, \times \nonumber \\
  & & \times \sum_i^{n-1}
  \frac{1}{2}[\sigma(M_i,z)+\sigma(M_{i+1},z)] \times \\ 
  & & \times \int_{M_i}^{M_{i+1}} \frac{\d
  n(M,z)}{\d M}\d M \ , \nonumber
\label{equation:tau}
\end{eqnarray}
where $n$ is the total number of clusters in our sample. This
corresponds to attaching to each halo of mass $M$ in the interval
$[M_i,M_{i+1}]$ a lensing cross section $\sigma(M,z)=1/2
[\sigma(M_i,z)+\sigma(M_{i+1},z)]$.

The integrand in Eq.~7 is shown in the left panel of
Fig.~\ref{fig:optd} for different cosmologies. The differential
optical depth for arcs with length-to-width ratio larger than $7.5$
and for sources at redshift $z_{\rm s}=1$ is larger in those
cosmological models where lenses form earlier and are thus more
concentrated. Moreover, the contribution to the total optical depth
comes from clusters in a wider redshift range in these
cosmologies. For example, in the high redshift tail, the curves drop
to zero at $z\sim0.65$ and $z\sim0.8$ in the $\Lambda$CDM model and in
the OCDM models, respectively. Moreover, at $z\sim0.6$, the
differential optical depth is still close to its maximum in the OCDM
model, while it is decreased below $30\%$ in the $\Lambda$CDM. Other
cosmologies, like the RP, the DECDM and the SUGRA with cluster
abundance normalisation, interpolate between these models, while we
obtain substantially smaller optical depths by adopting the CMB
normalisation of the power spectrum.

The total optical depth as a function of the normalisation of the
power spectrum is shown in the right panel of
Fig.~\ref{fig:optd}. Again, we find that when cluster normalisation is
used, dark energy models interpolate between $\Lambda$CDM and
OCDM. When the normalisation is reduced according to the amplitude of
large-scale temperature fluctuations in the CMB, the optical depth
also decreases as shown by the solid lines in the figure.

The total optical depth changes by a factor of two between the
$\Lambda$CDM and the OCDM models. This is in disagreement with the
previous results of \citet{BA98.2}, where one order of magnitude
larger optical depths were found in the OCDM compared to the
$\Lambda$CDM. However, we use here modified method for estimating the
optical depth. Moreover, both the mass and in particular the time
resolutions of our simulations are considerably larger than in the
earlier paper. This allows us to better account for the contribution
to the optical depth from all the merger events occurring in the
lenses, all of which enhance the cluster efficiency for producing
arcs, as shown in the previous sections. The impact of mergers appears
to be different in different cosmological models, being larger in
those cosmologies where lenses are less concentrated. As a consequence
of that, the difference in the number of arcs which is expected to be
observed in a $\Lambda$CDM and in a OCDM cosmological models is
reduced when mergers are accounted for. Indeed, resampling our cross
sections for reproducing the same time resolution of \citet{BA98.2},
we obtain an optical depth larger by a factor of eight in the
$\Lambda$CDM compared to the OCDM model, in agreement with
\citep{BA98.2}.

\begin{figure}
  \includegraphics[width=\hsize]{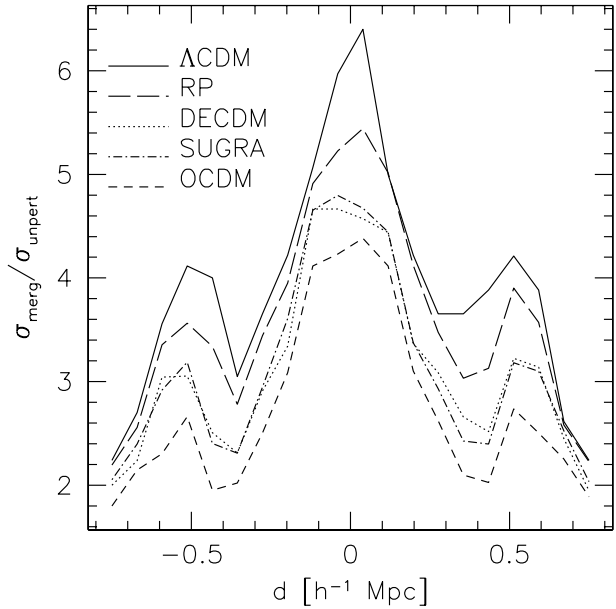}
\caption{Cross sections for arcs with $L/W>7.5$ as a function of the
  distance between two merging halos, modelled using a
  pseudo-elliptical NFW lens model with ellipticity $e=0.3$. The first halo has mass $M_{\rm
  main}=7.5\times 10^{14} h^{-1}M_{\odot}$, the second $M_{\rm
  sub}=3\times 10^{14} h^{-1}M_{\odot}$. The lens and the source
  planes are placed at $z_{\rm l}=0.3$ and $z_{\rm s}=2$,
  respectively. Different line types refer to different cosmological
  models.}
\label{fig:mergev}
\end{figure}

This interpretation is also supported by the following simple
experiment. We simulate the passage of a clump of matter of mass
$M_{\rm sub}=3 \times 10^{14} h^{-1}M_{\odot}$ through the centre of
an halo of mass $M_{\rm main}=7.5 \times 10^{14}
h^{-1}M_{\odot}$. Both the main cluster clump and the merging
substructure are modelled using the pseudo-elliptical
Navarro-Frenk-White lens model discussed by \citet{ME03.1}, assuming
an ellipticity of the lensing-potential contours of $e=0.3$. The lens
and source redshifts are kept fixed at $z_{\rm l}=0.3$ and $z_{\rm
s}=2$, respectively. Having placed the main halo at the centre of the
lens plane, we produce deflection-angle maps for several distances
between the merging clumps by moving the substructure along the
$x$-axis. Then, we apply the methods discussed in the previous
sections for determining how the lensing cross sections evolve as
functions of the distance between the colliding halos. Results are
shown in Fig.~\ref{fig:mergev}. Different line types refer to
different cosmological models and all curves are normalised to the
cross section of an unperturbed halo of mass $=7.5 \times 10^{14}
h^{-1}M_{\odot}$.

We note that in this analytic experiment we reproduce well the three
peaks in the lensing cross sections which were previously found in
numerical lensing simulations by \citet{TO04.1}. As discussed there,
the first and third peaks are due to the increasing shear in the
region between the two merging halos, while the substructure
approaches the main cluster clump. The second peak is due to the
larger convergence of the lens when the two clumps are perfectly
aligned. As anticipated earlier, the growth of the lensing cross
sections relative to the unperturbed case is larger by more than
$50\%$ in the $\Lambda$CDM than in the OCDM model. Other cosmological
models with static and dynamical dark energy interpolate between
$\Lambda$CDM and OCDM, confirming our assertion that mergers have a
stronger impact in cosmological models where halos are less
concentrated.

\citet{BA03.1} already estimated using analytic models the change of
the optical depth relative to the $\Lambda$CDM model in different
cosmological models with constant equation of state of the dark
energy.  The only cosmological model we investigated here, which can
be directly compared to the results of \citet{BA03.1} is the DECDM
model with CMB normalisation. In this particular case, we find an
optical depth which is larger by roughly $25\%$ compared to the
$\Lambda$CDM case, while the analytic models predicted a larger
increment, i.e.~between $50\%$ and $90\%$. This inconsistency is again
due to the smaller impact of mergers in the DECDM compared to the
$\Lambda$CDM model.

\begin{figure}
  \includegraphics[width=\hsize]{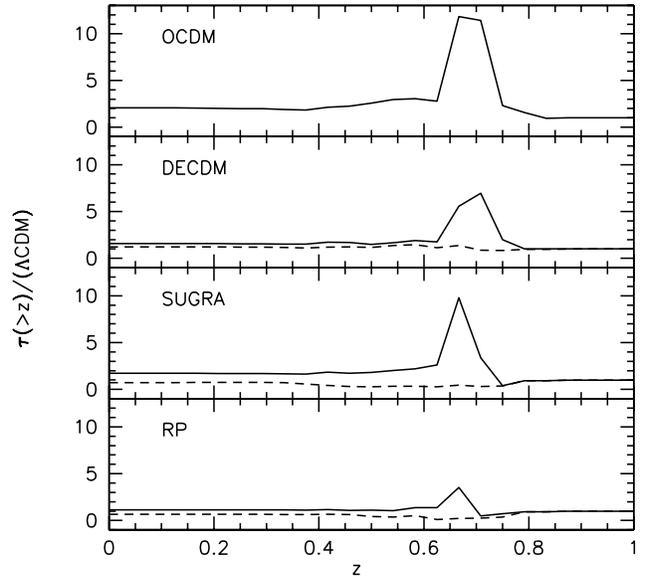}
\caption{Optical depth as a function of the minimal lens redshift
  contributing to the integral in Eq.~7. All cross
  sections of lenses below redshift $z$ were neglected.}
\label{fig:optdint}
\end{figure}

As discussed earlier, we see larger differences in the differential
optical depth among different cosmological models at high
redshift. This is better shown in Fig.~\ref{fig:optdint}, where we
plot the integrated optical depth as a function of redshift, when the
contribution from lenses below redshift $z$ is neglected. The curves
are normalised to the $\Lambda$CDM model. The curves show that the
deviations from the $\Lambda$CDM model become significantly larger at
redshifts higher than $\sim0.6$. Therefore, searching for arcs in
high-redshift clusters can be very promising for constraining both the
cluster evolution and the equation of state of dark energy.

\section{Summary}
We have investigated the efficiency of numerically simulated galaxy
clusters for producing strong lensing events in different cosmological
 models with constant and time-variable equation of state of the dark
energy, and we have compared it with ``standard'' $\Lambda$CDM and
OCDM models. Both the local abundance of galaxy clusters and the
observed amplitude of the temperature fluctuations in the CMB have
been used for normalising the power spectrum of the primordial density
perturbations generating our sample of numerically simulated
clusters. In the latter case, the normalisation is reduced for
accounting for the increasing ISW effect affecting in different ways
the large-scales CMB anisotropies in the cosmologies we
considered. When the local cluster-abundance normalisation is used,
the initial conditions in the N-body simulations were set up such as
to obtain halos having approximately the same evolution at redshift
zero.

Our main results can be summarised as follows:

\begin{itemize}

\item the lensing cross sections for long and thin arcs averaged over
  all clusters in our sample reflect the differences in the
  concentration of dark matter halos in different cosmological
  models. Halos are more efficient for producing strong gravitational
  lensing events in those cosmological models where they are more
  concentrated. Assuming the cluster abundance normalisation of the
  power spectrum, we find lensing cross sections which are smaller by
  roughly a factor of four in the $\Lambda$CDM compared to the OCDM
  model, and the other dark energy models interpolate between
  them. Moreover, despite their present equation of state is the same
  for the RP and the SUGRA models, their lensing cross sections differ
  substantially. For those dark energy models where CMB normalisation
  is adopted, we find lensing cross sections smaller by more than one
  order of magnitude compared to the OCDM model, indicating that the
  power spectrum normalisation is a crucial aspect for assessing the
  relative behaviour of the strong-lensing effect in different dark
  energy cosmologies.

\item We find that the local maxima in the lensing efficiency are
  caused by merger events occurring in the lenses. In particular, we
  verify that the peaks in the redshift evolution of lensing cross
  sections of individual clusters are shifted in time due to the
  offset between the merger histories in different cosmologies.

\item The optical depth for lensing is larger in those cosmological
  models where lenses form earlier and are thus more
  concentrated. Moreover, the contribution to the total optical depth
  comes from clusters in a wider redshift range in these cosmologies.

\item Compared to the $\Lambda$CDM model, the optical depth in the
  OCDM model is larger by only a factor of two, in disagreement with
  the results of previous other work, which found order-of-magnitude
  differences between these two models. However, the higher time
  resolution of our simulations allows us to account for the merger
  events occurring in the lenses, which we verified to have a stronger
  impact in those cosmological models where lenses are less
  concentrated.

\item When the same normalisation of the power spectrum is adopted, we
  find that clusters in the dark-energy models we investigated have
  optical depths within the boundaries set by the $\Lambda$CDM and the
  OCDM models. However, when the CMB normalisation is used, the
  optical depths become significantly smaller. The differences between
  the cosmological models are relatively small, but they become larger
  when the contribution to the lensing optical depths from high
  redshift clusters is considered.

\end{itemize}

Based on these results, we conclude that arc statistics is a
potentially very powerful tool for constraining the equation of state
of dark energy, and for investigating the dynamical evolution of
galaxy clusters. In particular, strong lensing events at high redshift
represent an important source of information which might be used for
discriminating between dark energy models. Further investigations are
needed now for fully understanding the impact of mergers on the
lensing cross sections and its dependence on cosmology.

\acknowledgements{The lensing simulations were carried out on the IBM-SP4
  machine at the ``Rechenzentrum Garching der Max-Planck-Gesellschaft
  und des IPP'' (Garching). The N-body simulations were performed at
  the ``Centro Interuniversitario del Nord-Est per il Calcolo
  Elettronico'' (CINECA, Bologna), with CPU time assigned under an
  INAF-CINECA grant. K.~Dolag acknowledges support by a Marie Curie
  Fellowship of the European Community program ``Human Potential under
  contract number MCFI-2001-01227.}

\bibliography{arcStatDe}
\bibliographystyle{aa}

\end{document}